\documentstyle[aps,pra,,twocolumn,epsf]{revtex}
\input{epsf}
\preprint{}
\begin{document} 
\title{Distribution of the delay time and the dwell time for wave reflection
from a long random potential}
\author{S. Anantha Ramakrishna\footnote{Presently at the Blackett Laboratory, Imperial College, London SW7 2BZ}  and N. Kumar\footnote{E-mail: nkumar@rri.res.in}}
\address{Raman Research Institute, C.V. Raman Avenue, Bangalore 560 080, India}
\maketitle
\begin{abstract}
We re-examine and correct an earlier derivation of the distribution of the 
Wigner phase delay time for wave reflection from a long one-dimensional 
disordered conductor treated in the continuum limit. We then numerically 
compare the distributions of the Wigner phase delay time and the dwell time, 
the latter being obtained by the use of an infinitesimal imaginary potential 
as a clock, 
and investigate the effects of strong disorder and a periodic (discrete) 
lattice background. 
We find that the two distributions coincide even for strong disorder, but
only for energies well away from the band-edges. 
\end{abstract}

\section{Introduction} The delay time associated with potential scattering is one of the important quantities related efficiently
to the dynamical aspect of scattering in quantum mechanics.  One of the common
measures for this quantity is the Wigner phase ($\phi$) delay time [$T_{\phi} =
\hbar (\partial \phi /\partial E)$]\cite{wigner}, which essentially entails
following a fiducial feature such as the peak of the wavepacket as it traverses
the scattering region.  This procedure is, however, rendered meaningless under
conditions of strong distortion of the wavepacket by the scattering potential
\cite{buttiker82,buttiker90}.  Further, there is the problem of indentifying the
position of the particle with the peak of the wavepacket.  Several researchers
have made other proposals for identifying a physically {\it meaningful}
timescale of interaction of the particle with the scattering potential (See for
recent reviews \cite{HandS,landauer94}).  These include the quantum clocks that
utilize the co-evolution, in a locally applied infinitesimal field / potential,
of an extra degree of freedom (such as the spin \cite{buttiker83}) attached to
the traversing particle.  Even these proposals are not completely free from
problems \cite{HandS,landauer94,golub90,sarsojourn}.  One related quantity that
has, however, remained uncontroversial is the dwell time obtained with the
`non-Unitary' clock \cite{buttiker90,golub90,sarsojourn,sardelay} involving
absorption/amplification due to a locally applied infinitesimal imaginary
potential ($V_{i}$)
\begin{equation}
\tau_{d}^{R} = \frac{\hbar}{2} \lim_{V_{i} \rightarrow 0}
\frac{\partial \ln \vert R \vert^{2}}{\partial V_{i}}, 
\end{equation}
for the case of total reflection ($\vert R \vert ^{2} = 1$).  In this case, it
also turns out to be the average Smith dwell time $\tau_{d} = (1/j) \int | \psi
|^{2} dx $, where $j$ is the incoming flux in the steady state situation, and
$\psi$ is the wavefunction in the scattering region of interest.  In this paper,
we will consider the Wigner phase delay time and the dwell time, given by the
non-Unitary clock, for total reflection from a long one-dimensional disordered
medium.  These times are, however, not self-averaging and one must have their
full probability distribution over a statistical ensemble of random samples.

The distribution of these times for the random media has been  investigated 
recently by several workers 
\cite{sardelay,jayanna,heinrichs90,gasparian,joshi1,joshi2,joshi3,texier,ossipov,vantiggelen,beenakker99,beenakker00a,schomerus00,heinrichs01}.
A delay time distribution that appears universal for wave reflection from a long 
one-dimensional (one-channel) random system was derived recently by Texier and Comtet\cite{texier} in the limit of high energy($E$) and weak disorder as 
\begin{equation}
\label{P0infty}
P^{0}_{\infty}( \tau ) =  \frac{\alpha}{\tau^{2}} \exp \left(- \frac{\alpha}
{\tau} \right) ,
\end{equation}
where $\alpha = 4 (\Delta^{2} k)^{-1}$, $\Delta^{2}$ is the strength of the
disorder (see Eq.~(6)), and the dimensionless delay time $\tau = E T_{\phi} /
\hbar $.  This was later confirmed by Ossipov {\it et al.}\cite{ossipov} for a
discrete random chain.

Earlier, we had derived the distribution of the dwell time for total reflection,
{\it i.e.}, in the insulating limit, 
using the non-Unitary clock for both passive as well as active (absorbing or
amplifying) one-dimensional random continuous media \cite{sardelay}.  The dwell
time distribution obtained by us, under the condition of a random phase
approximation which is valid for high energy and weak scattering, coincided
exactly with the delay time distribution of Texier and Comtet.  However, an
earlier calculation by Jayannavar {\it et al.}  \cite{jayanna} for the Wigner
phase delay time had obtained a slightly different form from the distribution
and it has been speculated \cite{ossipov} that this discrepancy may well be due
to the continuum model used by them.  In this paper, we first re-examine and
correct the calculation of Jayannavar {\it et al.}  for the Wigner phase delay
time distribution.  We find that the discrepancy noted above arises actually
from an inconsistency of the approximations made within the random phase
approximation (RPA), and that when the approximations are carried out
consistently, their expression reduces to the universal distribution as in
Eq.(\ref{P0infty}).  We then examine numerically the distribution of the delay
times and the dwell times for strong disorder using the transfer matrix method
for a one-dimensional disordered chain with a one-band tight binding
Hamiltonian.  We find that the distribution of the Wigner phase delay time and
the dwell time, clocked by the non-Unitary clock, agree exactly even in the
strong disorder regime, but for energies far away from the band-edge.  We
further examine the effect of a periodic lattice on the delay time by varying
the energy within the band.  We find that for strong disorder, and for energies
close to a band-edge, the Wigner phase delay time distribution differs
considerably from that of the dwell time given by the Non-Unitary clock.  The
Wigner phase delay time can even become negative under such conditions.  The
dwell time, however, remains positive as it must for total reflection ($|R|$ =
1).

\section{Distribution of the Wigner phase delay time for total reflection in the RPA}
Here we will re-examine the earlier calculation of Jayannavar {\it et al.}
\cite{jayanna,joshi1} for the distribution of the Wigner phase delay time for
total reflection from a one-dimensional (one-channel) disordered continuum. Again,
we will begin with the invariant imbedding equation for the reflection
amplitude $R(L) = \sqrt{r(L)} \exp [i\phi (L)]$ 
given by (in the notations of Ref.\cite{jayanna})
\begin{equation}
\label{reflect}
\frac{dR(L)}{dL} = 2 i k R(L) + \frac{ik}{2} \eta_{r}(L) \left[ 1+ R(L)
\right]^ {2} ,
\end{equation}
where $\eta_{r}(L) = -V_{r}(L)/E$ is the normalized fluctuating potential. 
In the limit of large lengths ($L \gg l_{c}$, the localization length), the
reflection becomes approximately total ($r(L) \simeq 1$), and Eq.(\ref{reflect})
yields an equation for the reflection phase $\phi (L)$ as
\begin{equation}
\frac{d \phi}{d L} = 2 k + k \eta_{r}(L) (1 + \cos \phi) .
\end{equation}
The equation for the phase delay time $T_{\phi} = \hbar (d\phi /dE) =
1/c_{g} (d\phi /dk)$ (where $c_{g}$ is the group velocity), is obtained by
differentiating the above equation for $\phi$ with respect to $k$:
\begin{equation}
\frac{d T_{\phi}}{dL} = \frac{1}{c_{g}} \left[ 2 + \eta_{r}(L) \left(
1 + \cos \phi - kc_{g} T_{\phi}\sin \phi \right) \right].
\end{equation}
As before, we will assume the random refractive index $\eta_{r}(L)$ to
be a Gaussian white noise with a zero mean, i.e.,  
\begin{equation} 
\langle \eta_{r}(L) \rangle = 0,~~~\langle \eta_{r}(L) \eta_{r}(L') \rangle = \Delta^{2} \delta (L-L'). 
\end{equation}
Using the Novikov theorem \cite{novikov}, we can now
set up a \ Fokker-Planck equation for the joint probability distribution
function $P(T_{\phi},\phi; L)$ over the ensemble of $\eta_{r}(L)$. However, we
will be interested in the marginal probability distribution $P(T_{\phi};L)$, 
which can be obtained by integrating over the phase angle $\phi$. To
this end, we make the random phase approximation (RPA) and set $P(T_{\phi},
\phi;L) = P(T_{\phi};L)/2\pi$, i.e., assume a factored out uniform 
distribution over the phase angle $\phi$.
The RPA is a good approximation for high energy and weak disorder\cite{rpa}. We obtain the equation for $P(T_{\phi};L)$ as
\begin{equation}
\label{wigtaueqn}
\frac{\partial P}{\partial l} = \frac{\partial}{\partial T_{\phi}} \left[
\frac{\partial}{\partial T_{\phi}} \left( \frac{T_{\phi}^{2}}{2} + \frac{3}
{2c_{g}^{2}k^{2}} \right) + \left( T_{\phi} - \frac{4}{c_{g} \Delta^{2} k^{2}}
\right) \right] P,
\end{equation}
where the dimensionless length $l = L/l_{c} = 1/2~ \Delta^{2} k^{2} L$. In the
limit of large lengths, $l \gg 1$, the distribution saturates and we can set
$\partial P/\partial l = 0$. Hence, we obtain the solution \cite{jayanna}
\begin{equation}
\label{wigtaudist}
P_{\infty}(\tau_{1}) = \frac{\lambda e^{\lambda \tan^{-1} \tau_{1}}}{(e^{\lambda\pi /2}
-1) (1+ \tau_{1}^{2})}\,\,,
\end{equation}
where $\lambda = 8/\sqrt{3} \Delta^{2} k$ and the dimensionless  time $\tau_{1} = c_{g} k T_{\phi} /\sqrt{3} $. This expression does yield the  
$\tau_{1}^{-2}$ {\em universal tail} behaviour
for $\tau_{1} \rightarrow \infty$, but differs from the distribution of
dwell times given by Eq.(\ref{P0infty}) at short times.  
The main difference appears at
$\tau = 0$, where this expression for $P_{\infty}(\tau)$ yields a finite value
in contrast to Eq.(\ref{P0infty}) which gives $P_{\infty}^{0}(\tau)=0$ because 
of the essential singularity at $\tau$=0.

This difference is readily traced to the fact that the RPA is good only in
the high energy,  weak disorder limit. Indeed, if we consistently take the high
energy, weak disorder limit in Eq.~(\ref{wigtaueqn}) [or in Eq.~(\ref{wigtaudist})], i.e., by demanding $c_{g}^{2}k^{2}
\rightarrow \infty$ and $\Delta^{2}/c_{g} \rightarrow 0$ with the
product $ (\Delta^{2}/c_{g})(c_{g}^{2}k^{2}) = 4/\alpha$ a constant,  we
obtain the solution as
\begin{equation}
P_{\infty} (\tau) = \frac{\alpha}{\tau^{2}} \exp ( -\frac{\alpha}{\tau} ),
\end{equation}
where $\tau = ET_{\phi}/\hbar$.
This is exactly the full universal distribution of delay times obtained  in Eq.~(\ref{P0infty})~\cite{texier} for the case of a free electron with $c_{g} = \hbar k /m$,
thus reconfirming again the universal delay time distribution. We note that the
above approximations have to be carried out consistently specially for a large
group velocity $c_{g}$, which is the case for energies far away from the band
edges. This also suggests that the condition of weak disorder $\Delta^2 k 
\ll$ 1 for the one-parameter scaling, which assumes a uniform distribution of the phase (RPA), may have to be modified to $\Delta^2k/(c_g/c_\phi) \ll$ 1, where $c_\phi$ is the phase velocity. 

\section{Strong disorder and a periodic background: Numerical
results}
The probability distribution of dwell times  in Ref.\cite{sardelay} 
was derived for a continuum model in the limit of weak disorder and high 
energy when the RPA is valid. In this Section, we will examine these 
limitations numerically. In particular, we investigate the distributions of 
dwell times for the case of strong disorder and compare the distributions of 
the dwell times and the Wigner phase delay times.  We will simulate a 
disordered lattice, instead of a continuum. The underlying lattice will also 
provide a discrete periodic background in the system, as distinct from a 
uniform continuum, whose effect  on the delay times will be investigated.

In order to go beyond the RPA, we will use the transfer matrix method 
involving the products of random transfer matrices\cite{transfermatrix}  to simulate the one-dimensional random medium using the one-band tight binding Hamiltonian (the Anderson Hamiltonian) with diagonal disorder \cite{economou}. 
The Hamiltonian describing the motion of a particle on the random lattice can be written as
\begin{equation}
{\cal H} = \sum_{n} \left[ \epsilon_{n} \vert n \rangle ~\langle n \vert  ~+~
V (\vert n \rangle ~\langle n+1 \vert +  \vert n+1 \rangle ~\langle n \vert)
\right]
\end{equation}
 where $\vert n \rangle$, $\epsilon_{n}$ and $V$ denote, respectively, the 
non-degenerate
Wannier orbital at the $n^{th}$ site, the site energy at the $n^{th}$ site and
the hopping matrix element connecting the nearest neighbours separated by a unit
lattice spacing. The site energies $\epsilon_{n}$ can be written explicitly in
the form of $\epsilon_{n} - i \eta$, with the real parts of the site
energies assumed to be independent random variables distributed uniformly over the
range $[-W/2,W/2]$ for $1<n<N$ and zero otherwise.
This is so that the disordered chain of $N$ sites is imbedded in an otherwise 
infinitely long ordered lattice. The imaginary part in the site energy 
($-i\eta$) makes the Hamiltonian non-Hermitian
and causes the particles to be formally coherently absorbed or emitted depending on the sign of
$\eta$, which is taken to be constant and non-zero (though infinitesimally small) over the disordered
segment $1<n<N$  and zero elsewhere. Since all the energies can be scaled with
respect to $V$, we will set $V$ to unity. 

The reflection ($R$) and the transmission($T$) amplitudes can now be calculated
using the transfer matrix method\cite{transfermatrix}. In order to calculate
the Wigner phase delay time, the reflection amplitude $R(E) =
\sqrt{r(E)} \exp[-i\phi(E)]$ is computed at two slightly differing values of
the incident wave energy, $E=E_{0}$ and $E = E_{0} + \delta E$, for a conservative
chain ($\eta = 0 $).  The Wigner phase delay time is then calculated as $T_\phi =
\hbar (d \phi/dE)= \hbar [\phi(E_{0}+\delta E) - \phi(E_{0})]/\delta E$.
Similarly, to calculate the dwell time by applying the imaginary potential,
the reflection amplitude is computed at two values of the
imaginary site energy ($\eta =0$ and $\eta = \delta \eta$). Now the dwell time
is given by $\tau_{d} = \hbar/2 (d \vert R \vert^{2}/ d \eta) = \hbar /2 [
\vert R(E,\eta=\delta \eta)\vert^{2} - \vert R(E,\eta=0) \vert^{2}]/\delta
\eta$. Typically the values of $\delta E$ and $\delta \eta$ are $10^{-6}$ and
the stability of the results have been checked for their choice within the range
$10^{-5} < \delta E,~\delta\eta < 10^{-7}$. (We will deal with the delay dwell 
time in a dimensionless form by multiplying it by $V$ and setting $\hbar=1$.) For the
calculation of the averages and the distributions, we have typically used
$10^{5}$ configurations of the disorder. We will present our results for a long 
sample  ($L\gg l_{c}$, i.e., lengths much greater than a localization length).

We will first examine the case of wave energies far away
from a band edge ($E=0.0,~1.0$). In Fig. \ref{midbdelay}, we show the
distribution of the Wigner phase delay time $\tau_{w}$ and the dwell time
$\tau_{d}$ for reflection from a long sample for different values of the
disorder strengths ($W = 0.1,~2$). For weak disorder ($W =0.1$), the
distributions are identical to each other and also correspond exactly to the
universal distribution given by Eq.(\ref{P0infty}).  It should be noted that the RPA is not valid for exactly the band centre ($E=0$) due to a well-known anomoly, although it is valid for a generic value of energy within the band\cite{anomoly}. The two distributions
also coincide for higher disorder strengths.\cite{footnote}
It is interesting to note that Eq.(\ref{P0infty}) still describes 
the distribution reasonably well for moderately large
disorder ($W=2.0$, see Fig. \ref{midbdelay}b),
though the RPA under which the expression was derived is not valid for these
cases. The case of $E=1$ shows similar behaviour, though the peak occurs at a
different value, reflecting the smaller group velocity.  In Fig.~\ref{midbdelay}d, we plot the distributions of dwell and delay times for cases of symmetric disorder (the random site energy is chosen from the range symmetric about zero $[-W/2,W/2]$) and asymmetric disorder (the site energy can only  be positive $[0,W]$ or negative $[-W,0]$). The distributions for the positive and the negative one-sided, asymmetric disorder
appear to be the same, regardless of the sign as expected, of course (we have 
included these two only as a check on our numerics). 
These, however, are different
from the distribution for the symmetric case. The contribution of the prompt
part of the reflection arising from the average potential mismatch at the
boundary is clearly seen for the cases of asymmetric disorder in that the
peak of the distribution  occurs slightly earlier, and there is more weight 
at early times. The asymmetric disorder model changes on an average the lattice potential locally only over the disordered segment (sample) and {\it not} 
globally over the sample and the leads. This gives rise to a potential 
mismatch at the ends of the disordered chain and cannot be absorbed by a 
mere shift of the incident energy. More formally, in the case of the  symmetric
 disorder the average of the S-matrix, $\langle S\rangle =0$, while in the case of the asymmetric disorder, $\langle S \rangle \neq 0$.
%

Now, we will examine the case of wave energies close to the edge of the band
($E=1.9,~1.99$). In Fig. (\ref{edgebdelay}), we show distributions of the delay
time and dwell times.
For the case of weak disorder, again the Wigner delay
time distributions and the dwell time distribution coincide. There is, however,
considerable discrepancy from Eq.(\ref{P0infty}), as can be seen. We  have 
explicitly verified that the RPA is valid for this   case of weak disorder by
calculating the distribution of the phase. Thus, the discrepancy cannot be an artefact of the RPA. The most probable reason, perhaps, is that near the band edge, the wave does not penetrate deep enough to fully sample the randomness, before getting reflected promptly. This would call into question the factorization of the joint probability distribution of the phase and its energy derivative, particularly for short times. 

For intermediate and strong disorder ($W =1,~2,~3$), a more interesting effect
occurs.  The two distributions, i.e., the Wigner delay time distribution and the
dwell time distributions no longer coincide.  The difference between the
distributions increase with the disorder strength and with their proximity to
the band-edge.  The Wigner delay time distribution appears quite different from
the universal distribution at $E=0$.  Near the band-edge, in fact, for $E=1.99$ and $W=1$, the
Wigner delay time distribution is non-zero for even negative times.  This is,
presumably, due to the strong deformation of the wavepacket caused by the strong
dispersion near the band-edge.  The dwell time distribution given by the
`non-Unitary' clock, however,  remains non-zero only for positive times.  
We also note that
the Universal $\tau^{-2}$ tail at long times ($\tau \rightarrow \infty$) remains
unaffected.

\section{Conclusions}

In conclusion, we have studied the distribution of the delay and the dwell times
for reflection from a disordered medium in the limit of total reflection ($|R|$ = 1).  We have revisited the original calculation of
Jayannavar {\it et al.}\cite{jayanna} for the distribution of the Wigner delay
time.  We show that, by taking the high-energy limit consistently within the
RPA, the correct universal distribution of delay times is reproduced.  In the
course of the derivation, we note that the single-parameter scaling ansatz for
the RPA seems consistent under the condition $\Delta^2k/(c_g/c_\phi) \ll 1$
($\Delta^2$ -the disorder strength, $c_g$-the group velocity and $c_\phi$-the
phase velocity), instead of $\Delta^2k \ll$ 1 which does not account for the
effects of the group velocity.  This is in accord with the recent results of
Ref.  \cite{ossipov,altshuler00}.  We have also investigated the distribution of
delay times numerically and find the distributions of the Wigner delay time and
the dwell time to coincide for energies far away from a band-edge for all
disorder strengths.  This, however, breaks down for energies close to the
band-edge and strong disorder, when the dispersive effects of the band structure
deform the wavepacket so much so as to render the description in terms 
of the motion of a
wavepacket meaningless.  The concept of a dwell time, clocked by a counter such
as the imaginary potential, however, remains meaningful even under such
circumstances.

\section*{Figure Captions}

1. The distribution of delay and dwell times for reflection from a long
disordered passive medium for
wave energy at the middle of the band ($E=0.0, ~1.0$). \\

2. The distribution of delay and dwell times for reflection from a long
disordered passive medium for
wave energy close to the edge of the band ($E=1.9, ~1.99$).

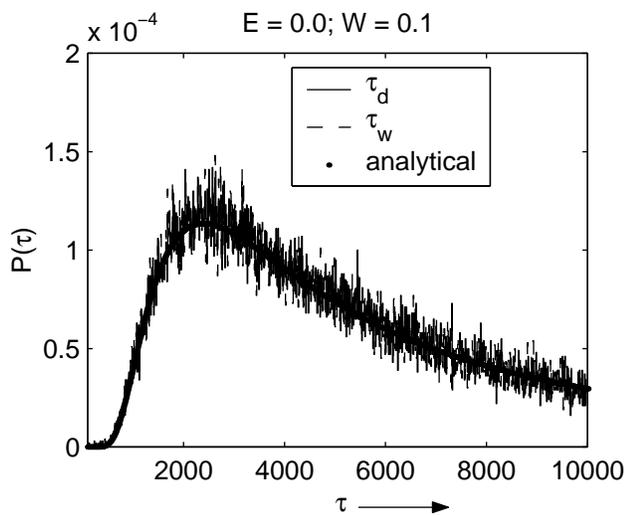
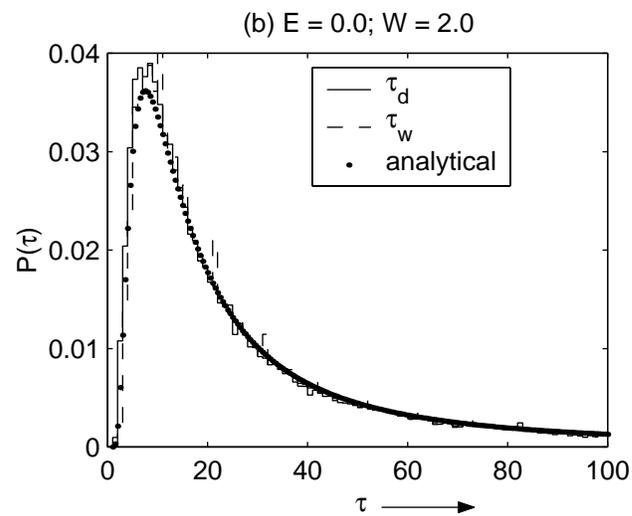
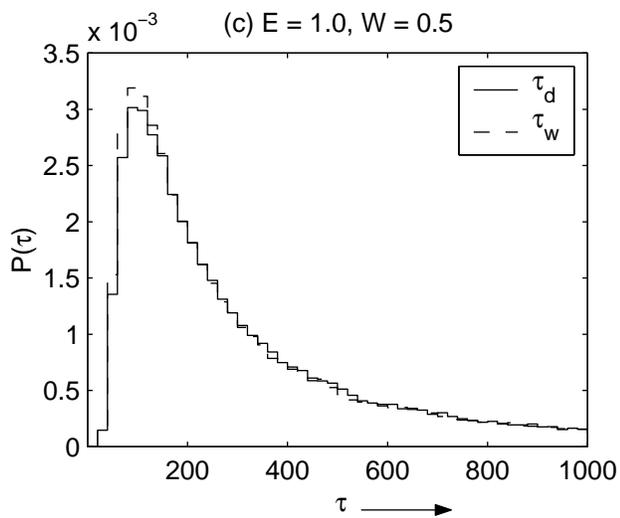
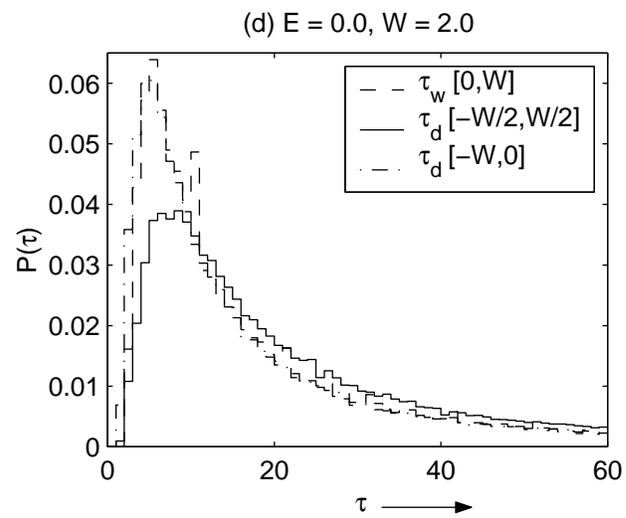
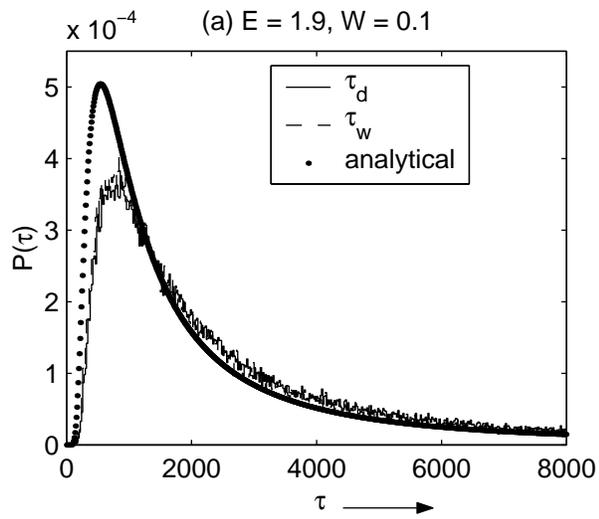
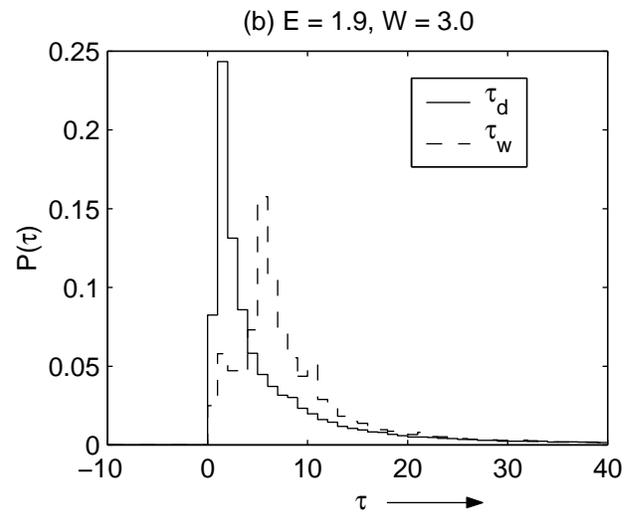
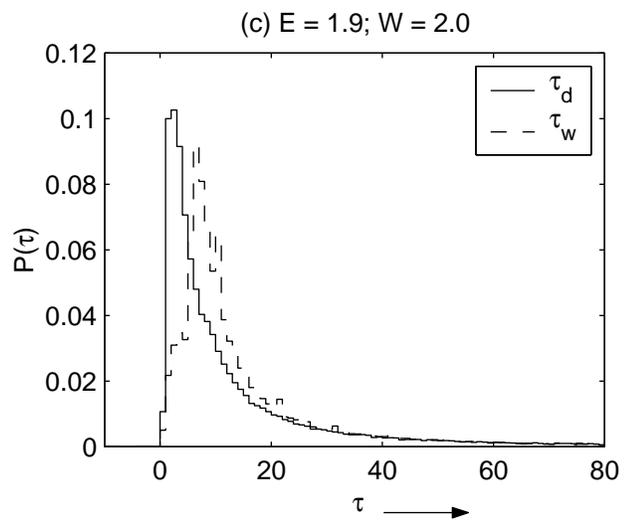
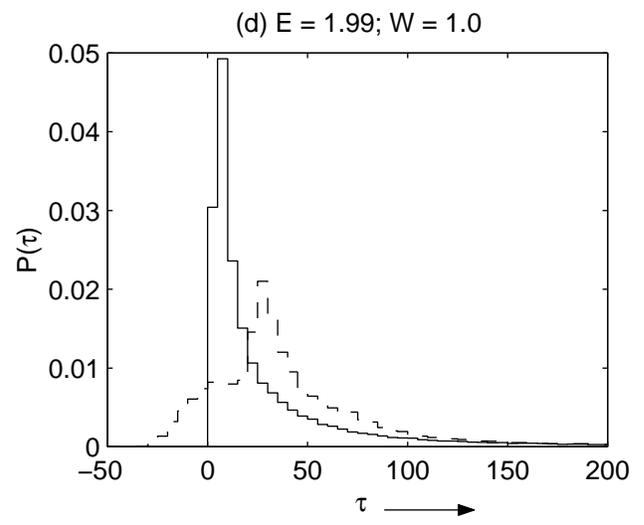
\end{document}